# Improving Privacy and Trust in Federated Identity Using SAML with Hash Based Encryption Algorithm

Jissy Ann George
College of Administrative and Financial Services
AMA International University
Kingdom of Bahrain
jissy.george@amaiu.edu.bh

S.Veni
Department of Computer Science
Karpagam University
Coimbatore, Tamil Nadu
venikarthik04@gmail.com

Safeeullah Soomro
College of Computer Studies
AMA International University
Kingdom of Bahrain
s.soomro@amaiu.edu.bh

*Abstract* - Cloud computing is an upcoming technology that has been designed for commercial needs. One of the major issues in cloud computing is the difficulty to manage federated identities and the trust between the user and the service providers. This paper focuses on how security can be provided between the user and the service provider and how the user information can be authenticated. For the purpose of providing privacy and authentication, Security Assertion Markup Language (SAML) based Single Sign-On is used. Security is provided by using Hash based Encryption algorithm (HBE). HBE algorithm works with the help of Key Exchange Protocol which contains poly hash function. In the algorithm, Identity providers maintain user directory and authenticates user information; service provider provides the service to users. The user has to register their details with the identity provider prior to this. During this stage, Hash based Encryption algorithm is used to provide secure communication between the identity provider and the user. In this paper we suggest that higher security can be given to user login by using an additional cryptographic technique, i.e. Hash based Encryption algorithm with the help of the Key Exchange Protocol.

I. INTRODUCTION

Cloud computing provides different kinds of services to users with different charges based on the particular usage of services. In cloud, the services are cost effective and those services [1] are easy to use. But the main issues are access control, security, maintaining and managing the user identities. When providing service to the user, their identities are needed to manage high security, and for this purpose, identity as a service (IDaaS) is used. IDaaS is very important in all services including software, platform, and infrastructure service in both public and private clouds. Therefore, identity management system is very important because in the enterprise environment, the application data may be interrupted by a third party. At the same time, issues such as consistency in authentication, authorization and auditing are handled by identity management using the identity layer which was in IDaaS.

IDaaS manages all user information using different identity management models like isolated, central, centric and federated. Providing security to user identities and managing trust between the user and the service provider are the main issues. In the proposed system, the identities are managed by using Security Assertion Markup Language with Single Sign-On [2] and encryption algorithm. SAML is an extensible markup language which is used for user communication to the service provider with the help of particular authentication [3]. The single sign-on method provides adequate security for user entities in the identity management, hash based encryption algorithm is used. At first the user has to register the details in identity provider and they must encrypt the user details with a particular length of key which is used to manage the user information for further processing. During the sign-on processing, complete assertion and authentication process helps to increase privacy of service related identities. In HBE, the user has a different number of keys which is used to validate the password with encryption and decryption keys. The user identifier, a 128-bit keyword is used to provide complete security and authentication to access service from the cloud. The key has to be changed using a different internet key exchanging protocol because it only provides additional security to user information and integrity check. Key exchange protocol has used particular hash functions to authenticate message code while transferring secret keys in the cloud. The key protocol first encrypts the user information using a secret key and then it is transmitted to further security based service access. Encryption algorithm provides the best security because both the service provider and the user have different encryption and decryption keys while sharing the password through the internet. So while an identity provider hides the user attributes from being accessed by a third party. In addition, it is easy to identify the third parties who are trying to access the user information using the encryption and decryption keys with different protocols. Additionally, this proposed system covers the lack of service access, process and data management using the encryption algorithm. The following section describes the identity management related works, and security related proposed system.

## II. RELATED WORK

Cloud computing is one of the main resources for developing a new business with a minimum cost and flexible platforms. It has provided different kinds of services but the main challenges are privacy, security, authentication, authorization and access control. Therefore, user information management, trust between the user and the cloud service providers have to be managed using different kinds of security and encryption algorithms. Federated identity management is the main feature to manage the user information using different security concepts like OpenID, OAuth, SAML with Single Sign on method. Eghbal Ghazizadeh et al., [4] proposed using trusted computing, federated identity management and OpenID Web SSO to solve identity theft in the cloud. They mention that identity issues are managed by using OpenID but some of the attacks lead to low security in identity management. They further mention that OpenID requires further steps to control the identity theft. Cloud computing is one of the main technologies for providing different kind of services to the service requestor. The service provider needs to maintain security to manage user information and identity management.

Hongwei Li et al. [5] proposed different identity based cryptography techniques to be used to manage security. Hierarchical based cloud computing architecture is built and then security is managed by using identity based encryption, identity based signature with an authentication protocol like Secure Socket Layer (SSL) to manage user side information. Finally, the security in cloud computing is established using different cryptography techniques. Cloud computing is a developing scenario for new generation which provides different kinds of resources to service requestors.

Nida et al., [6] focused on the importance of Identity and Access Management (IAM). IAM is identity and resource management within the organization and it is the building block for the information security program and the most widely interacted security areas by the user. Identity and access management is, therefore, one of the most important concepts which helps to manage remote access user credential. In cloud computing, security is the main issue but security changes from one cloud identity model to another. Unique identification and authentication are important when providing services to the cloud service requestor.

Bernd Zwattendorfer et al., [7] identified different numbers of cloud identity management models that are available like isolated, user centric, central, federated and federated broker model. The user information has to be managed, based on the identity model. They proposed that federated broker based identity model should provide the best security and authentication to the user information. Identity based management system provides some security issues while dealing with the federated identity model.

Liang Yan et al., [8] showed the need to combine WS-Security approach federated identity management and HIBC and how it can reduce security problems in the hybrid cloud. This method is advantageous compared to other security algorithms and it reduces the SOAP header size. They have also shown that identity based problems can be restricted by using Hybrid Identity Based Cryptography (HIBC).

Roshni Bhandari et al., [9] discussed different identity management techniques for providing authentication, authorization, non-reputation, data confidentiality. In this paper author discusses about the various identity management frameworks such as SAML, OpenID, OAuth, PRIME, OneLogin and identifies related usage areas.

Jan Vossaert et al., [10] explained that user centric based identities are managed by using different trusted models. The user centric based model overcomes the federated identity model problem. It also has the extra functionality to provide security and trust between the service provider and the user. In the proposed system, the author demonstrated flexibility and more secure properties.

Md. Sadek Ferdous [11] implemented an approach which does not require change in SAML. It focuses on trust issue which is managed by dynamic federation. SAML based identity federation is created by users. They provided a means of creating dynamic federations automatically. Antonio Celesti [12] showed how trusted inter-domain communication is established using CLEVER based cloud. Authentication and trust between different CLEVER domain is achieved by SAML based SSO profile. It obtains authentication, trust between users and the cloud provider using SAML and CLEVER based cloud.

## III. SAML AND HBE FOR IDENTITY MANAGEMENT

In cloud computing identity management is the main issue, because a large number of users requests the enterprise for various services. At the time identities are managed by using different federated identity protocol like OpenID, OAuth, SAML [13]. In the existing systems, the services are accepted by using the user name and the password but the user credential is hacked by using different phishing attacks. So, it is difficult to maintain user credential using the Single Sign-On (SSO) method and the main drawback of the existing system is the trust between the user and service provider. Another problem is the identity provider and the service provider may misuse the user information during authentication and authorization. Therefore, in federated based identity model, it is very difficult to manage the user information when several users request the service.

Federated based management system also focuses only on the Personal Identification Information (PII) during that time and is thus difficult to provide security to user credential.

**Proposed System:** In federated identity model, it is difficult to manage multiple user credential details for authentication and authorization data between parties in spite of managing user identities from different security and authentication issues. In the proposed system, the user identities are managed by using the User-Centric identity management model with Security Assertion Markup Language based Single Sign-On algorithm for providing authentication between the user and the identity provider because the user centric model works based on Personal Trusted Device (PTD) with the help of Personal Transaction Protocol (PTP). In the identity model, the user information and the credential details are stored in the identity provider so that the user information is stored in the Personal Authentication Device (PAD) [14]. PAD is the context of the computer security which provides the particular key like PIN number to the user, so the user can access the different number of services using a single PIN. In addition, SAML provides secure login with Personal Authentication Device so that the user can store in unlimited details in the service provider with a single sign-on login. This procedure enables authentication and security to identities. Another problem is trust between the user and service provider. This drawback can be overcome by using encryption algorithm, Hash based Encryption (HBE). The user password is encrypted and decrypted using this encryption algorithm with key exchange protocol with poly hash function. The service provider and the user use different 128 bit key for encryption and decryption, so no one can access the user identities. Finally, encryption and decryption process increases the trust between the user and the service provider.

**Security assertion and markup language:** Security Assertion and Markup Language is one of the Extensible Markup languages. This is used to provide authentication between the user and the identity provider with the help of different protocols like HTTP, SOAP, and XML. SAML has three different components, namely assertion, binding, protocols, which give security to the user credentials [15]. These components are used between the Identity provider (IP) and the Service provider (SP) which indicates how it works together with the single sign-on. The single sign-on method is initiated by the identity provider or the service provider. If the service is initiated by IP, assertion is signed, encrypted or both. The browser or the user requests the resource from IP using their user name and password; then the IP redirects the authentication request to the browser. The browser then gets the authentication for the username and the password using a particular encryption algorithm and posts the request to the service provider. After that, SP gives resources to users. Figure 1 shows Single Sign-On authentication via SAML [16].

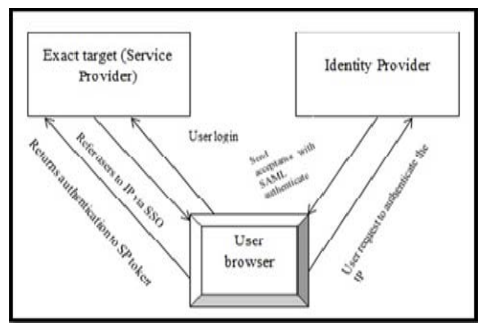

Fig. 1 User Authentication using Single Sign-On via SAML

**Hash based encryption algorithm:** In cloud computing, communication is done with the help of media data so the data can be prevented from phishing and malicious attacks. For protecting the data, cryptography techniques are used. In the proposed system, Hash Based Encryption cryptography technique is used because it overcomes several problems like security and attacks. In HBE, fixed 128 bit of plaintext is used for encryption with different key lengths of 128 bit, 192 bit, 256 bits. In cloud, the services are accessed from different unknown service providers so the user identities need to be saved for further processing. Authentication and authorization is maintained by using Key Exchange Protocol (KEP). Poly message authentication code is combined with protocol which increases the security while transferring the user password with the known public key and secure private keys. Following steps describe the procedure of HBE algorithm.

**Step by Step Procedure of HBE Algorithm**
- HBE processes the entire data block and in parallel each round perform substitution and permutation.
- Input has 128bit and input related key is expanded into forty-four 32bit words.
- It has four different stages to provide the security one for permutation and the other three for substitution.

1. **Substitute Bytes**
   Different block ciphers used for special substitution called 'S-box' which is used to perform byte by byte substitution.
2. **Shift Rows**
   A simple shifting is performed row by row, but the row 0 is never changed.
3. **Mix columns**
   Substitution of the alerts of each byte in a column as a function of all the bytes in a column
4. **AddRoundKey**
   Simple X-OR operation should be performed between the current block and the expanded key.

For both encryption and decryption cipher text started with the AddRoundKey and followed by nine rounds with each performing those four stages.

The proposed system SAML with HBE encryption algorithm and key exchange protocol provides a high level security for the user identity management. Figure 2 explains the basic work flow between the user, the Identity Provider and the Service Provider using Security Assertion Markup Language with Hash based Encryption algorithm.

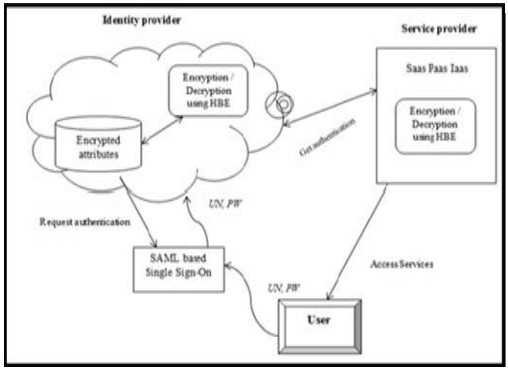

Fig. 2 SAML and HBE for Identity Management

At first, the user requests the identity provider to access the particular service from the cloud using his own user name and password. Here the identity management is managed using the user centric identity model so that users need not memorize all the details. He or she just remembers that electronic PIN number. After that, the Identity Provider accepts the user request and authenticates the user information and returns the acceptance message to the user with help of SSO login via SAML [17]. Here the security and user information is managed by a third party, but in the cloud trust between user, identity provider and service provider is one of the main issues. For overcoming this issue, encryption algorithm with fixed input 128 bit block text and variable key length encryption algorithm is used to authenticate the user information.

It is proposed to use the hash based encryption algorithm, because the key length is changed every time and the key expansion plays an important role in security and authentication system. HBE algorithm work is also based on nine rounds and followed by the tenth round. For every round it has four different stages like substitution bytes, shift rows, mix column, key expansion. For every round the key length has to be increased (e.g. 128 bit has 10 rounds, 192 bit has 12 rounds, and 256 bit has 14 rounds). The secret key has to be exchanged between the user and the service provider with poly hash function. Thus encryption and decryption between the identity provider and the service provider helps to avoid the misuse of user information.

## IV. RESULTS AND DISCUSSION

In this proposed work, implementation is done using Java platform. SAML is used for implementing both service provider and identity provider communication with Single Sign-On (SSO). To address the security issue, cryptographic technique Key Exchange Protocol (KEP) – Hash based Encryption Algorithm is used to ensure secure communication. Table 1 shows the average time taken to run HBE algorithm with different key lengths.

TABLE I
HBE RUNNING TIME WITH DIFFERENT KEY LENGTH

| HBE with different key | Megabyte processed | Time taken | MB/second |
|---|---|---|---|
| 128 | 256 | 2.976 | 44.386 |
| 192 | 256 | 3.196 | 41.010 |
| 256 | 256 | 3.817 | 33.145 |

So, it has been shown that HBE algorithm provides better security in cloud because it takes several billion years to crack the 128 bit key using a brute force attack. Table 2 displays the time to taken to crack key versus and key size.

TABLE II
TIME TO CRACK CRYPTOGRAPHIC KEY VERSUS SIZE

| Key size | Time to crack |
|---|---|
| 128 bit key | 256 |
| 192 bit key | 256 |
| 256 bit key | 256 |

From Table 2, it is easy to know that phishing and brute force attacks cannot hack the user information because it takes more time and it is also difficult to find the secret key from the cloud. SAML with HBE algorithm has a minimum computation time, which is clearly explained using Table 3 and Figure 3.

TABLE III
EXECUTION TIME TAKEN FOR ENCRYPTION

| Encryption without hash function | | Encryption with hash function | | SAML with HBE | |
|---|---|---|---|---|---|
| Key Size | Execution time(ms) | Key Size | Execution time(ms) | Key Size | Execution time(ms) |
| 128 | 3.567 | 128 | 2.9 | 128 | 1.7 |
| 192 | 4.985 | 192 | 3.9 | 192 | 2.4 |
| 256 | 6.126 | 256 | 5.2 | 256 | 3.3 |

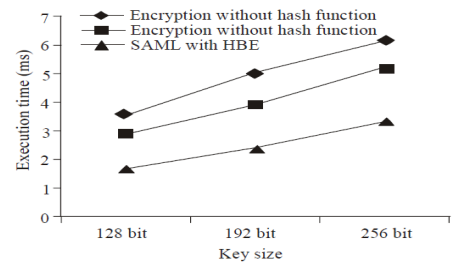

Fig. 3 Time taken for computing Encryption

Figure 3 shows the time taken for computing Encryption with and without hash function. During the key schedule, it allocates different number of rounds and those round related secret keys work with key exchange protocol.

Hash based key encryption combined with SAML results in less encryption time. The algorithm works with poly function,

which increases the security between user identities. So, HBE algorithm has been proved to provide the best security and also build the trust between the user and the service provider with a minimum execution time.

## CONCLUSION

In this paper, cloud security and trust between user and the Service Provider have been provided by using Security Assertion Markup Language with Single Sign-On. Even though it provides authentication to user login, by using an additional cryptographic technique, i.e. Hash based Encryption algorithm, high security with the help of the Key Exchange Protocol can be obtained. That the identities and user credential details are managed by user centric identity management model has been thus proved through implementation in this paper. Thus, different enterprises request for and receive their services through cloud using SAML user web browser with a high security, and a fast acceptance.

## ACKNOWLEDGMENT

Part of this work was done for my Thesis work Titled, "Improved User-Centric Identity Management Solution Using Cryptographic Techniques for Cloud Security".